\documentclass[sigconf]{acmart}

\AtBeginDocument{%
  \providecommand\BibTeX{{%
    \normalfont B\kern-0.5em{\scshape i\kern-0.25em b}\kern-0.8em\TeX}}}

\setcopyright{none}

\acmConference[ICSE 2022]{The 44th International Conference on Software Engineering}{May 21–29, 2022}{Pittsburgh, PA, USA}

\usepackage{color}
\usepackage{nameref}
\usepackage{multirow}
\usepackage{tabularx} 
\usepackage{censor}
\usepackage{nameref}

\begin{document}

\title{A Grounded Theory of Coordination in Remote-First and Hybrid Software Teams}


\author{Ronnie E. de Souza Santos}
\affiliation{%
  \institution{Faculty of Computer Science}
  \institution{Dalhousie University}
  \city{Halifax}
  \state{NS}
  \country{Canada}
}
\email{souzasantos.ronnie@dal.ca}

\author{Paul Ralph}
\affiliation{%
  \institution{Faculty of Computer Science}
  \institution{Dalhousie University}
  \city{Halifax}
  \state{NS}
  \country{Canada}
}
\email{paulralph@dal.ca}

\begin{abstract}
While the long-term effects of the COVID-19 pandemic on software professionals and organizations are difficult to predict, it seems likely that working from home, remote-first teams, distributed teams, and hybrid (part-remote/part-office) teams will be more common. It is therefore important to investigate the challenges that software teams and organizations face with new remote and hybrid work. Consequently, this paper reports a year-long, participant-observation, constructivist grounded theory study investigating the impact of working from home on software development. This study resulted in a theory of software team coordination. Briefly, shifting from in-office to at-home work fundamentally altered coordination within software teams. While group cohesion and more effective communication appear protective, coordination is undermined by distrust, parenting and communication bricolage. Poor coordination leads to numerous problems including misunderstandings, help requests, lower job satisfaction among team members, and more ill-defined tasks. These problems, in turn, reduce overall project success and prompt professionals to alter their software development processes (in this case, from Scrum to Kanban). Our findings suggest that software organizations with many remote employees can improve performance by encouraging greater engagement within teams and supporting employees with family and childcare responsibilities. 

\end{abstract}


\maketitle

\section{Introduction} \label{sec:introduction}
Since the beginning of the COVID-19 pandemic, remote work has become the reality for millions of professionals around the globe~\cite{brynjolfsson2020covid,collins2021covid,craig2021dual}. In fact, the number of professionals that will be permanently working from home is expected to continuously increase \cite{phillips2020working} \cite{chavez2020permanently}. 

Software development and IT professionals, in particular, are more likely to embrace fully or partially working from home, for at least two reasons: (1) much infrastructure for supporting working from home has already been developed for open source, distributed, and global software engineering projects, and is common in the industry; (2) many of these professionals will be involved in creating and adapting technologies to support work-from-home (and other remote and hybrid part-remote/part-office arrangements) for employees in other industries.

Much research has already investigated distributed and global software development in the sense of outsourcing arrangements between on-shore and off-shore partners, or teams spread across multiple offices in different countries. Much research has also investigated the benefits and drawbacks of working from home and other forms of teleworking in many different industries. The situation we investigate is different in several ways: co-located teams accustomed to continuous face-to-face interaction were forced into working-from-home whether they wanted to or not, on short notice, with little preparation, during a worldwide crisis. It differs from global software engineering because we are not looking at an onshore organization outsourcing to offshore partners; and differs most telework literature in that employees did not elect to work from home for personal reasons. 

Moreover, working from home during a pandemic differs significantly from working from home in normal circumstances \cite{ralph2020pandemic,ford2021tale}. Many workers found themselves in improvised home-offices, awkwardly sharing space with family members, or juggling work with childcare (as day cares closed) or managing children newly enmeshed in messy online replacements for regular school. 

And yet, many software professionals have grown accustomed to the flexibility and astonishingly short commute times, and want to keep working from home all or part of the time. Many anecdotes have emerged of developers threatening to quit if required to return to the office \cite{Frishberg2021These}. 

Therefore, while the effects of the COVID-19 pandemic may be temporary (we hope), studding this intersection of adversity and working from home is important to understand how long-term trends toward more working from home, remote-first teams, distributed teams, and hybrid teams may affect the software industry. This raises the following research question. 



\smallskip
{\narrower \noindent \textit{\textbf{Research question:} How has working from home due to the COVID-19 pandemic affected software professionals and software teams?} \par}
\smallskip

To address this question, we conducted a year-long, participant-observation, constructivist Grounded Theory study at a large, South American software engineering company. As is common in Grounded Theory, we begin with the preceding high-level research question and allow more specific questions and answers to emerge from constant comparison and theoretical sampling. 

During this study, we found it necessary to define the following work arrangements.
\begin{itemize}
    \item In a \textit{co-located} team, all team members work in the same physical space (e.g. the same building) most or all of the time. A single team member might work remotely for a few days (e.g. while feeling unwell), or the whole team might work remotely for a day (e.g. when the office lost power) but the \textit{default} work arrangement is everyone in the same physical space. 
    \item In a \textit{distributed} team, team members are spread across two or more office spaces in different geographic locations (typically different cities; sometimes different countries). Team members may visit each other, but the \textit{default} work arrangement is multiple offices in different locations.
    \item In a \textit{remote-first} team, the default work arrangement is for each team member to work in their own work space (e.g. a home-office, a coffee shop, a desk in a co-working space). The team may or may not have centralized office space.
    \item In a \textit{hybrid} team, on any given day, some team members may be working in a co-located office space while others are working remotely. Hybrid teams can result from some team members always working remotely, from all team members sometimes working remotely, or some combination thereof. 
\end{itemize}

Next, we discuss existing studies about team coordination and about the effects of the pandemic in software industry (Section \ref{sec:background}). In Section \ref{sec:method}, we describe how we conducted the study, while Section \ref{sec:findings} presents our findings. In Section \ref{sec:disussion}, we discuss the implications and limitations of our study. Finally, Section \ref{sec:conclusion} summarizes the contributions of this study. 

\section{Background} \label{sec:background}
This section presents related works, divided into two categories: research on coordination and research about how COVID-19 affected software engineering. 

\subsection{Team Coordination}
The study of team coordination is not new in software engineering, especially because the software industry became accustomed to projects that are developed on a global scale, with professionals scattered across different locales \cite{herbsleb2007global}. Consequently, software professionals became more adaptable towards working in remote and distributed environments. 

``Coordination means integrating or linking together different parts of an [organization, group, team, community, etc.] to accomplish a collective set of tasks'' \cite[p. 322]{van1976determinants}. In software engineering, team coordination is the process of managing dependency among activities of different professionals in the team \cite{herbsleb2007global}. Coordination has been reported as a key decision-making factor \cite{herbsleb2003formulation}, and the management of such decisions directly impact project success. Coordinating a software project means accommodating the perspectives of different software professionals about their tasks, and how they are related to their collective goal \cite {van1976determinants}; that is, project success. 

Previous research has established that coordination mechanisms are shaped by two dimensions of communication, namely, vertical coordination and horizontal coordination \cite{nidumolu1995effect}. \textit{Vertical coordination} refers to a coordination process based on the hierarchic structure established within the project (e.g. a project manager deciding which team member will perform which tasks). \textit{Horizontal coordination}, in contrast, refers to professionals organizing amongst themselves, establish mutual agreements and adjusting as necessary (e.g.  programmers selecting their own tasks) \cite{nidumolu1995effect}. 

Although coordination theories \cite {herbsleb2003formulation} suggest that communication is the key to coordination, practitioners struggle to determine what is effective coordination in their teams, or what is a good level of communication and how to enhance it. This happens because coordination problems go beyond communication. Usually they involve other project aspects, such as, number of professionals involved in the project, time to perform tasks, technical matters, and task complexity \cite{herbsleb2006collaboration}. 

Moreover, in software engineering, co-located projects and distributed projects will present their own particularities in relation to team coordination \cite{herbsleb2007global}. In summary, software projects vary in their need for coordination either in the amount of coordination required by the team, or in the type of coordination depending on the context (e.g. co-located, distributed) \cite{herbsleb2006collaboration}.


\subsection{COVID-19 and Software Engineering}

Initial investigations of the effect of COVID-19 on software engineering found that working from home during the pandemic had negatively affected professionals’ wellbeing and productivity, and that these effects may be worse for women, parents, and people with disabilities \cite{ralph2020pandemic}. Many individuals struggled to adapt to their new routine and had issues with connectivity, distractions, and time management \cite{ford2021tale}. In addition, the pandemic triggered new managerial challenges and communication issues \cite{butler2021challenges}. Meanwhile, the possibility that the pandemic will lead to a permanent increase in working from home may create both advantages and problems to software professionals, especially those working in large projects \cite{bao2020does}.  

However, one longitudinal study found that developers adapted to lockdown challenges over time \cite{russo2021predictors}. Moreover, some evidence suggests that agile practices can be adapted successfully to overcome some of the limitations of the remote work; for instance, adaptations on the team meetings might be required to improve communication strategies and developing knowledge sharing approaches \cite{da2020agile}. Furthermore, working from home---even in these less desirable circumstances---has some benefits including proximity to family and increased flexibility, which appear to increase individual motivation \cite{ford2021tale,bezerra2020human}. 
These effects can vary across contexts; for instance, while proximity to family is beneficial for some professionals, it can also undermine work-life boundaries and cause distractions, interruptions \cite{ralph2020pandemic,ford2021tale}. 


Finally, some studies highlighted the software industry's role in supporting society. Research focused on mining software repositories demonstrated that open-source technologies can be applied rapidly to deal with worldwide emergencies, since many projects covering various aspects from COVID-19 have been developed using open-source technologies, which were not affected by the lockdowns \cite{wang2020open}. On the other hand, the global crisis triggered intense activity on Stack Overflow\footnote{\url{https://stackoverflow.com/}}, with professionals interacting on topics related to the analysis of COVID-19 data \cite{georgiou2020preliminary}. Such findings reinforce the ability of software professionals to collaborate remotely to create software-based solutions. 


\section{Method} \label{sec:method}

This section describes our research method and site. Briefly, we used Constructivist Grounded Theory \cite{charmaz2014constructing}, emphasizing participant observation, to study a software development organization as it coped with pandemic-induced disruptions. We began with a broad topic---the impact of the pandemic on software professionals and their projects---and iteratively collected and analyzed data to generate a theory. The more specific research question stated in Section \ref{sec:introduction} emerged simultaneously with our core category. 


Grounded Theory is a family of research methods for inductively generating theory. It is characterized by specific techniques including interleaved rounds of data collection and analysis, inductive coding, memoing, constant comparison, and theoretical sampling \cite{glaser1978theoretical},\cite{charmaz2014constructing}. It focuses on investigating real-life settings to identify concepts that explain behaviours and experiences. While many variants of Grounded Theory exist, we employ \textit{Constructivist} Grounded Theory because it is epistemologically consistent with our perspective on qualitative research; namely, that the researcher constructs rather than discovers knowledge. Moreover, we blend data collection approaches common to Grounded Theory (e.g. interviews, document analysis) with approaches more often associated with ethnography (i.e. participant observation), because doing so allows greater triangulation and insight.



\subsection{The Site} \label{sec:site}
The site is Recife Center for Advanced Studies and Systems (CESAR), a well-established, mature software company in South America. The company was founded in 1996 and specializes in on-demand software development for external clients. It applies advanced engineering in information and communication technologies to solve complex problems for national and international companies and industries from various sectors including finance, telecommunication, manufacturing, and services. It has over 900 employees of which about 70\% work directly on software development teams (i.e. as programmers, QAs, designers, analysts, etc.). The first author began working for CESAR as a quality assurance analyst in February 2020. 

At the time of the study, these professionals were working on approximately 50 different projects, running simultaneously, and applying a wide variety of software processes from Waterfall to Scrum \cite{schwaber2002agile} to the Rational Unified Process~\cite{kruchten2004rational}. Some projects have many developers with no quality assurance analysts (because an external client is doing their own quality assurance), while others have only quality assurance with no developers (in which the client was responsible for the code and the company responsible for testing it), and many in between.

When the pandemic escalated in Brazil, the company immediately applied emergency measures to protect its professionals and keep the business running. These measures affected at least 900 employees directly and over 3,000 clients and users indirectly. Some important early events were as follows: 

\begin{itemize}
\item \textit {Early March 2020:} professionals were informed that a committee was created to study the pandemic scenario, establish measures, and take action regarding the crisis.

\item \textit {March 12:} the committee determined that anyone with chronic diseases, under health treatments, or who lives with someone with chronic disease should start working from home immediately.

\item \textit {March 15:} the committee determined that every employee, team, and project ready to work remotely should do so immediately, while those who are not ready should be remote within three days.

\item \textit {March 16--19:} the company kept professionals updated about the progress of moving to working from home.

\item \textit {March 20:} all professionals working in software development activities were working from home. 
\end{itemize}

We selected the site and topic of the study simultaneously. When the company, shifted to remote work, the first author noticed many interesting changes and ramifications. In early September 2020, the first author described the research opportunity to the second author, and we decided to pursue the study. We requested permission from the company and filed an ethics application with the second author's university. The ethics application was approved on November 6 and the company agreed on November 9. In summary, we did not ``select'' the site as much as identified an opportunity at a company where interesting things seemed to be happening. 

The first author participated as a quality assurance analyst in two projects during the study period. The first project involved developing and maintaining a mobile application for an international company. The project included two developers, two quality assurance analysts, a tech leader, and a manager. 

The second project involved developing a desktop application for the same client. The project included the same two quality assurance analysts, tech leader and manager, and four different developers. Both projects used Scrum. 

\subsection{Data Collection}

We applied three data collection techniques: participant observation, collecting asynchronous communications, and semi-structured interviews. Note that our \textit{dominant} data collection technique is participant observation. While most Grounded Theory studies are interview-centric, we used interviews and virtual communications to corroborate and elaborate concepts mainly emerging from participant observation. 

Data collection began on November 10, 2020 and ended in August 2021. The data included communications and artifacts from early March, 2020, to July 2021. The first author also recreated field notes from memory and documents (including email and Slack messages) going back to early March 2020. (While this entails a memory threat to validity, as we shall see below, the concepts arising from these reconstructions were all elaborated and corroborated through real-time observation and interviews.) 

\subsubsection{Participant Observation}

Our primary data collection mechanism was participant observation; that is, observing events while participating in them. Participant-observation allows researchers to take part in the routine of a targeted group for a continued period to observe actions, behaviours, interactions, and events related to a topic under investigation \cite{aktinson1998ethnography}. 

Being part of the project teams, the first author observed and participated in sprint planning meetings, daily scrum meetings, biweekly retrospective meetings, and regular meetings between the team and the client, and organization-wide meetings about company matters, such as the progress of the emergency measures for the pandemic, the company's situation in the face of the pandemic, and strategies for remote work and return to the office.

\subsubsection{Asynchronous Communications}

Meanwhile, we collected asynchronous communications including emails, discussions on the company's Slack,\footnote{\url{https://slack.com}} and retrospection boards (maps of concerns raised during retrospection meetings). We did not dragnet and download all of the teams' voluminous discussions; rather, we copied snippets that seemed relevant to the topic of the study.    

In both observing events and collecting asynchronous communications, we were influenced by the principals of netnography and digital ethnography. Netnography and digital ethnography are online adaptations of ethnographic methods. While netnography takes place in online environments by observing interactions among the members of online communities \cite{kozinets2002field}, digital ethnography focuses on virtual communication content, both graphic (e.g. images, emoticons, videos) and verbal (e.g. posts and discussions) \cite{kudaibergenova2019body}. Both strategies applied throughout the study, since field notes were created based on observed messages, posts and media about the pandemic, the emergency measures (i.e. in March, 2020) and employees interactions while working from home, through posts on the employees’ general channels (e.g. Slack and email lists). 

\subsubsection{Interviews}

We also interviewed not only members of the two observed project teams but also other company employees with different profiles and backgrounds, working on projects with different development methods, in different domains. We interviewed these additional participants to enrich our data, obtain additional evidence and assess resonance of our emerging theory. 

We conducted semi-structured interviews with 20 professionals between November 12, 2020 and August 20, 2021. Interviews were conducted through Slack following a pre-established interview guide. Interview time varied from 31 to 67 minutes (average of 43 minutes). We began by interviewing all members of the two project teams mentioned above except the manager (who was invited but declined due to workload). We then invited participants who we suspected could shed light on phenomena we had identified but did not totally understand yet (i.e. theoretical sampling). This produced 13 hours and 48 minutes of audio and 287 pages transcripts. 

Participants were asked about their experiences switching from working at the office to working from home, preparedness (theirs and the company's), advantages, challenges, impacts of working from home on their work activities and the project, lessons learned and expectations for the future. The exact questions varied depending on the interview and evolved in keeping with theoretical sampling. An example script is available (see \nameref{sec:DataAvailability}). 

\subsubsection{Engaging Participants}
Beyond interviews, we also engaged with participants by initiating discussions on their (Slack) channels about topics related to the research. This allowed us to refine our data and to improve our understanding of actions, behaviours, and nuances that we observed. Through this process, participants were frequently invited to share their opinions about events or topics that affected the whole company (e.g. decisions about returning to the office; internet connection problems). These limited interactions allowed for greater flexibility and more cross-checking than confining ourselves to longer, more structured interviews. 

\subsection{Data Analysis}
We began data analysis on field notes and asynchronous communications before the first interview. We began with line-by-line ope coding on our initial field notes to identify emerging concepts and their properties \cite{glaser1978theoretical}. To analyze non-text artifacts (e.g. images) we wrote memos describing them, and then treated the memos like field notes. 

Next, we began conducting interviews. Following each interview (or two, when they were quite close together) we transcribed the interview, applied open coding, and compared the new codes to our growing collection of codes (i.e. constant comparison). Audio recordings supported this process of giving meaning to the codes. 

Around the twelfth interview, we shifted more toward categorizing codes (i.e. focused coding). While the core category remained elusive, we establish links among different codes, based on an intense analysis focused on observing the categories and their interconnections.

By July 2021, we still had not settled on a core category. There were so many interesting aspects to the case that we had trouble choosing the most important thing to focus on. We decided to analyze the retrospection boards developed by the project teams during their biweekly retrospection meetings. Retros are often informative since they facilitate reflection on a cycle of work and highlight positive events as well problems faced by the teams. The first author had attended these meetings and recorded many of their events in field notes already, but going back over the issues raised during the retrospection meetings helped elevate the core problems and challenges faced by the teams. 

We analyzed the retro boards 
by integrating field notes from retros with discussion topics and action items lifted from the retro boards created by the teams. When we analyzed the retros, the core category became clear and the proposed theory rapidly emerged and saturated as the remaining important categories fell into place. 

Next, we began theoretical coding: iteratively rearranging our categories until they stabilized and confirming the connections built among them. Simultaneously, we solicited feedback on our findings from willing participants (i.e. member checking resulting in minor suggestions on our observations and concepts). We continued until we could clearly define each category and relationship in our emerging theory. The combination of these definitions, positive comments from member checking, and a shared perception that all of the main concepts had been included indicated that the theory was saturated.  

\begin{figure*}[ht]
  \centering
  \includegraphics[width=0.9\linewidth]{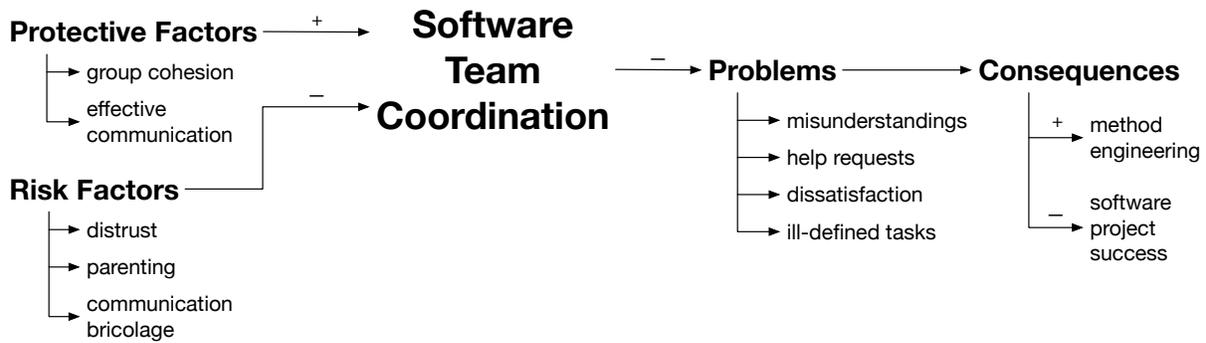}
  \caption{Causes and Consequences of Software Team Coordination}
  \Description{A nomological network showing how protective factors and risk factors affect coordination, the lack of which creates problems and consequences}
  \label{fig:theory}
\end{figure*}

All of our data was stored in an encrypted shared storage provided by our university. Coding was performed mostly in Microsoft Word and Excel. While we have tried more sophisticated coding tools (e.g. NVivo), we find word processing and spreadsheet tools are often easier to work with. We drew and iterated on the theory diagram (Figure \ref{fig:theory}, below) in OmniGraffle.\footnote{\url{https://www.omnigroup.com/omnigraffle}}  

\subsection{Auditing}

The bulk of the coding was performed by the first author. The interview transcripts, and therefore the initial coding, was all in Portuguese. As we moved toward theoretical coding, the first author translated relevant categories and quotations. The second author audited these later (English) coding stages, helping to clarify and elaborate the emerging theory. The auditing processes led to significant reorganization and clarification of the emerging theory (e.g. dividing coordination antecedents into protective factors vs. risk factors; distinguishing problems from consequences of those problems). 

\subsection{Ethics}
We generally followed the norms of the ethics committee at the second author's university, which approved this research. The company agreed to participate in this study, allowing data collection on general channels and email. In addition, each professional who was interviewed individually agreed in providing data, and all members of the observed project teams consented to participate. Before each interview, the interviewer explained the general goal of our research and its relevance for the software industry. Participants were guaranteed data confidentiality (except for publishing brief, translated quotations that, in our best judgment, would not identify them) and informed about the voluntary nature of the participation, along with the right to withdraw from the research at any moment. No participants withdrew from the research. 

Data collection began after ethical approval. Analysis of emails, posts, and discussions that occurred in channels was completed retroactively after ethics approval (with the consent of the participants involved), enabling the data collection process to obtain information from the initial transition to work-from-home in March 2020. Interviews included questions not only about the present but also reflecting on past experiences and expectations for the future.

\section{Findings} \label{sec:findings}

This section explains the categories that emerged from the analysis, and how they combine to form a theory of coordination in remote software teams. Figure \ref{fig:theory} illustrates the proposed theory; Table \ref{tab:definitions} briefly describes its concepts, which we further elucidate next. Table \ref{tab:chainOfEvidence} illustrates the chain of evidence from observations to theoretical concepts. 

\begin{table*}
  \caption{Brief Descriptions of Theory Concepts}
  \label{tab:definitions}

\begin{tabularx}{\linewidth}{l X} 
\toprule
Concept & Definition \\
\midrule
Communication bricolage & Professionals creating ad-hoc communication mechanisms for subgroups (rather than the entire team) \\
Dissatisfaction & Unhappiness resulting from experiencing a given situation  \\
Distrust & Feelings of suspicion regarding others' work \\
Effective communication & A state in which most information exchanges are successful \\
Group cohesion & ``The tendency for a group to stick together and united in the pursuit of its goals and objectives'' \cite[p. 572]{carron1982cohesiveness}  \\
Help requests & Increased need for assistance from others to complete tasks \\
Ill-defined tasks & Ambiguous work; team members perceiving the same work differently, or not perceiving a task as having a clear beginning, middle and end \\
Method engineering & Participants intentionally modifying their ways of working (e.g. from Scrum to Kanban) to mitigate coordination problems \\
Misunderstandings & Misinterpretation and mistakes regarding what needs to be done and what is actually accomplished \\
Parenting & Caring for one's children \\
Project success & The cumulative effects of a software engineering endeavor on its stakeholders over time \cite{ralph2014dimensions} \\
Software team coordination & The ability of individuals to collaborate effectively on their individual tasks to achieve a collective objective \\
Protective Factors & Phenomena that improve coordination, or mitigate coordination breakdowns \\
Risk Factors & Phenomena that hinder coordination or contribute to coordination breakdowns \\
Problems & Challenges caused \textit{directly} by coordination breakdowns \\
\bottomrule
\end{tabularx}

\end{table*}

\begin{table*}
  \caption{Example Evidence for Theory Concepts}
  \label{tab:chainOfEvidence}

\begin{tabularx}{\linewidth}{p{1.6cm} p{1.8cm} X}
\toprule
Theme & Category & Example Evidence \\
\midrule
Protective\newline factors & Group\newline cohesion & ``it's not like when you are at the same place with your project mates, you don't know how to do something, you discuss that problem with someone'' (P4) \\
& & ``there must be some practices that can be used to keep the team together and integrated'' (P1) \\
& & ``to get people more interested in meetings, thematic meetings were created, such as, the Tuesday-costume daily meeting and the English-only day'' (Field Notes [Feb 22, 2021]) \\
& Effective & ``they try to bring people close together, debate together, share information'' (P2) \\
& {communication} & ``we try to keep the documents always updated and a link available for easy access'' (P11) \\
& & ``we extended how we register the information, e.g., verbal versus written'' (P14) \\
\midrule
Risk\newline factors  & Distrust & ``you have always this need for transparency, because many people are always worried about 'how things are going'?'' (P5) \\
& & ``professionals are constantly meeting to obtain each others validation to the solutions they are implementing'' (Note taken observing Project 2) \\
& & ``but I feel that when you work at the office and when you know the person face to face, and when we are working together, near each other, we trust more in the work that someone is doing'' (P20) \\
& Parenting & ``we had meetings that were interrupted by kids playing around at home and you know that there is nothing to do'' (P5) \\
& & ``children will require time, they make noise as well, sometimes parents can't concentrate'' (P13) \\
& & ``for example, [NAME OF COLLEAGUE] who has two kids at home. So, for him, it's more difficult to focus than me believe'' (P20) \\
& Communication bricolage & ``many discussions happen between individuals using specific channels and not everybody in the team knows it'' (P11) \\
& & ``nowadays, for instance, I have to keep different tabs open with Whatsapp, Gatherm, Slack, MS Teams and Email to keep informed about what is going on'' (P14) \\
& & ``many synchronous meetings happening at the same time'' (Field notes [Oct 8, 2020]) \\
\midrule
Problems & Misunder- standings & ``we agreed with a change requested by the client, but [NAME OF COLLEAGUE] was not in the meeting and did not implemente what was required'' (P16) \\
 & & ``we had misunderstandings so we decided to document more details regarding the process, results, meetings and decisions'' (P17) \\ 
 & & ''during the retrospective the team decided that a checklist of requirements needs to be created to avoid misunderstandings among developers and QAs'' (Field notes [Jul 02, 2021]) \\ 
& Help requests & ``it's hard to conduct training activities, specially with new members'' (P13) \\
& & ``since I work with mobile devices and we use several gadgets, several times I have to ask someone to help me'' (P16) \\
& & ``young professionals require more support, and sometimes more help'' (P17) \\
& Dissatisfaction & ``people get upset when they are told they would have to go back to the office'' (P16) \\
& & ``I was working too much (...) and this was making me dissatisfied with the company'' (P03) \\
& & ''there is a noticeable dissatisfaction about the number and the duration of meetings to keep the team synchronized, which is sometimes keeping individuals to work on their normal hours.''(Field notes [Jun 11, 2021]) \\
& Ill-defined & ``sometimes you do more than what is your responsibility, aiming to help the team'' (P12) \\
& tasks & ``there was a lack on the definition of how the team would be managed'' (P16) \\
\midrule
Consequences & method   & ``we now have weekly meetings to discuss our team work ... so we can synchronize expectations'' (P12) \\
& engineering & ``we changed from a two week Sprint model to a Kanban model with continuous deliveries'' (P15) \\
& & ``we needed to change our process some times ... to identify priorities, impacts on our work and become more efficient'' (P16) \\
& & ``some tools and ceremonies needed to be included to look the team closely, such as 1:1 meetings'' (P18) \\
& project  & ``remote work made me work more hours. Overtime. This anticipated some project demands'' (P13) \\
& success & ``sometimes we need to delay releases'' (P16) \\
& & ``we kept delivering with the same quality [same as the office], but I believe we need to put more effort into it; in summary, we have increased the number of responsibilities'' (P17) \\
\bottomrule
\end{tabularx}

\end{table*}

Working from home affects software professionals in many ways. Similar to previous studies \cite[e.g.][]{ralph2020pandemic}, we found that individuals exhibit problems associated with ergonomics, preparedness, and well-being. However, the core category emerging from our data is \textit{coordination}. Specifically, our analysis reveals that working from home causes intra-team coordination problems that interfere with software development and trigger changes to software development processes.

\subsection{Coordination} \label{sec:coordination}
Coordination is an important part of software development because most software is created by groups (rather than individuals) and the activities of these groups are interdependent. Groups need (formal or informal) coordination mechanisms not only to distribute tasks but also to complete them without inhibiting each other (e.g. breaking each other's code) \cite{brooks1995mythical}. Poor coordination thus leads to duplicate work, rework, merge conflicts, delays and other forms of waste \cite{sedano2017software}. 

Coordination within and between software teams has been discussed extensively in the context of distributed and global software development \cite{herbsleb2003formulation,herbsleb2007global}. Distributing individuals across multiple geographic locations, especially different countries and time zones hinders coordination \cite{wagstrom2014does}.

However, our participants all worked from homes within the same city. We therefore observed different kinds of coordination problems and strategies for improving coordination (discussed below) from those commonly discussed in the context of global software development. For example, we did not observe coordination problems caused by cultural differences or time zone disparities between onshore and offshore teams. 

Before the pandemic, face-to-face communication was the main coordination mechanism at the site. Each team worked together, at the same site, and the company arranged the workspace in a way that everybody on a team was physically near to each other. Few formal coordination mechanisms (e.g. meetings, documentation) were needed. 

Once individuals began working from home, however, the cornerstone of their coordination was compromised. Professionals did not know for instance what time their teammates were available or when they could reach out for discussing tasks. The usual ``quick chat by my teammate's desk or workstation`` used to discuss project matters (suddenly) did not longer exist. That is, one of the most simple communication strategy was taken away by the new work arrangement. In addition, professionals were sharing their space with family members that were not exactly understanding the level of concentration required to conduct software development activities. Hence, teams were forced to search for alternatives to keep their work on sync. 

Teams rapidly adopted new coordination mechanisms; for instance, sprint planning and retrospective meetings. Teams also used various technologies to support coordination, including Slack for synchronous and asynchronous communication, reminders that were automatically triggered during the day (e.g., update your tasks status or join the daily meeting), and shared documents for decisions and notes. 

In summary, we found that working from home increased teams' need for coordination while simultaneously undermining their core coordination strategy (face-to-face discussion). This drove the teams to coordinate through more formal, consistent, technology-mediated synchronous communication (i.e. meetings) and asynchronous communication (e.g. shared documents; Slack messages).

\subsection{Protective Factors}
We found two factors---namely \textit{group cohesion} and \textit{effective communication}---that appear to improve coordination and cooperation among software professionals, shielding software teams from the negative effects of remote work on coordination.   

Group cohesion is ``a dynamic process which is reflected in the tendency for a group to stick together and united in the pursuit of its goals and objectives'' \cite[p. 572]{carron1982cohesiveness}. In other words, a cohesive team has shared goals and a sense of shared identity. Team members \textit{feel} like they belong to the team. They feel free to interact, and the synergy developed among them supports their cooperation. A large but inconsistent body of literature suggests that group cohesion is an important antecedent of performance in diverse domains \cite{casey2009sticking}.  

Our participants felt that group cohesion was strong when they worked together, on-site, before the pandemic. They felt that they had good, open relationships and interactions with their teammates. We observed that before the breakdown they were constantly discussing their activities, brainstorming, and bounding. As the teams moved to working from home, coordination was preserved to the extent that the team's cohesiveness was maintained. We observed that teams that were unable to keep individuals interacting in a regular basis started to face coordination problems as the individuals started experiencing detachment in relation to their teammates. 

Similarly, effective communication appears essential to preserving satisfactory coordination. Coordination requires communication because team members have to share information to complete interdependent tasks. For instance, we observed that everyone on the team needs to have the same level of understanding about a feature under development. However, requirements are a liability and can always change. When change happens, the understanding needs to be carried across the team, and poor communication will result in misinterpretation and lead to errors. In addition, we observed that effective communication will allow professionals to be synchronized on the progress of the activities towards achieving the team common goal. 

However, remote work prevents highly effective face-to-face communication among teammates, which means that strategies (i.e. virtual communication approaches) to increase effective communication are needed to support team coordination. In this sense, we observed teams having more meetings. The main objective of these meetings was getting everyone informed about the status of the activities, as well as understanding dependencies among activities. Before the transition to remote work, this constant syncing was accomplished through face-to-face, informal communication. We also observed teams attempting to make meetings more engaging using gimmicks (e.g. costume-wearing day).

This perspective differs from previous research, which categorized issues in global software development into communication, coordination and control \cite{aagerfalk2006old}. Rather than separate yet related kinds of issues, we observed a clear precedence wherein communication supports coordination; not the other way around. 

\subsection{Risk Factors}
We also identified several risk factors---namely \textit{distrust}, \textit{parenting}, and \textit{communication bricolage}---that can undermine coordination of software teams working remotely. Participants described these risks as challenges that require constant attention to avoid interfering with team coordination.

Working from home appears to decrease software professionals' trust in each others' work. When the team worked in co-located office space, team members tended to trust that their colleagues' work was executed in the right way. When the team shifted to working from home, participants reported feeling more suspicious of whether everyone is aligned towards the same objectives. Increased distrust led participants to spend more time and effort validating completed work. We observed that distrust negatively affected the interdependence among tasks and work autonomy, because individuals are always actively involved with each others tasks at some level creating a feeling of involuntary cooperation. Therefore, as distrust affects cooperation, it is a risk for team collaboration.

Meanwhile, our findings reinforce reports from previous studies that parenting became a major challenge during the pandemic \cite{ralph2020pandemic}. Caregivers, especially those responsible for small children, had to adjust their routines more than their childless colleagues to cope with competing work and household responsibilities. Our participants (both parents and childless co-workers of parents) indicated that having young children at home interfered with coordination in several ways, including: 
\begin{enumerate}
    \item children, and caring for them, can be distracting, which can negatively affect the integration with teammates;
    \item caregivers are more often unavailable for synchronous activities, which  increases reliance on asynchronous communication and coordination.  
\end{enumerate}


As explained above, effective communication appears to protect team coordination. However, 
we observed professionals creating ad hoc communication mechanisms, including individual Slack messages and emails, for sub-groups of their teams. Sharing information and discussing issues with sub-groups instead of the whole team created knowledge silos. As information is spread, replicated and modified, professionals struggled to synchronize and maintain shared goals. This represents a major risk for team coordination since it interferes on the the team unity (i.e. the team breaks down into several groups). We refer to this disintegration of team communication as \textit{communication bricolage} because the knowledge silos arise from team members making do with imperfect substitutes for regular face-to-face communication.

\subsection{Problems}
Above, we discussed protective factors that support coordination and risk factors that undermine coordination. On the opposite side of our model, we have problems \textit{caused by} poor coordination, and consequences of those problems. 

Ineffective team coordination is a serious problem for software development because it reduces productivity and delays delivery of new products, features, updates and fixes. Beyond these inherent effects, however, we found four additional problems associated with poor coordination of remote software teams: 

\begin{enumerate}
\item \textit {Misunderstandings}: team members have differing ideas of what needs to be accomplished and how (e.g. the correct behaviour of a system feature).

\item \textit {Help requests}: ineffective coordination increases requests for help among team members, either to solve problems, to understand activities, or to clarify the team's proximate goals. For instance, we observed professionals frequently asking for help to understand requirements and make decisions (e.g. what is the best way to implement a piece of code; what is essential to be tested), whereas face-to-face communication used to be enough.

\item \textit {Dissatisfaction}: participants indicated that they felt the need to work more due to decreasing productivity and increasing delays. They experienced increased guild and unhappiness when something was wrong with the project because they were at home and not in the office. This experience appears to drive a general decrease in participant's job satisfaction. 

\item \textit {Ill-defined tasks}: lack of coordination hampers professionals' ability to perceive the boundaries between their tasks and their colleagues' tasks. In other words, team members have trouble identifying the beginning, middle, and end of their tasks. We observed that because distrust drove a constant need to validate each other's work, professionals lost their sense of autonomy---they struggled to decide when they were done with a task or when they needed to collaborate with a colleague to actually finish it.

\end{enumerate}

Each of these problems can manifest simultaneously and may be interrelated. Moreover, they will depend on several factors including the background of the professional (i.e. year of experience, previous experience with remote work), the home office arrangement (i.e. separate and quiet place to work, sharing place with other people), and individual differences such as personality (i.e. more introverted, more focused, more into discussions, etc). However, exploring such factors is tangential to our main theme and therefore left to future work.  

\subsection{Consequences}
The four problems described above appear to hinder the overall success of the software project and encourage teams to adapt their software development process. 

We observed teams making several process adjustments to adapt to working from home. Some adjustments involved minor changes on the dynamics of ceremonies (e.g. time of the sprint planning and duration of the stand-up meetings). Other adjustments were more sweeping; for instance, both teams we observed mitigated coordination problems by switching from Scrum \cite{schwaber2002agile} to Kanban \cite{Brechner2015Kanban}. Specifically, they replaced two-week sprints with releasing small features as soon as they were completed. However, they kept the dynamics of the meetings, such as, planning, dailies and retrospectives. These teams felt that changing their software process was necessary to deal with delays in the releases, e.g., delivery on one specific date. 

Furthermore, team coordination appears closely related to project outcomes. We did \textbf{not} observe reductions in software quality (i.e. comparing remote work and working at the office) or major delivery delays during the period of this research. However, professionals reported that completing their activities took more time (including overtime) and effort. In other words, the teams completed their work on time and to the expected level of quality, but at greater cost to themselves. We model this as a negative effect on overall success following a stakeholder-impact view of success \cite[cf.][]{ralph2014dimensions}. 

\subsection{Summary} \label{sec:summary}

In summary, we found that shifting from in-office to at-home work fundamentally altered coordination within software teams. Group cohesion and effective communication appear to help preserve coordination; however, distrust, parenting and communication bricolage appear to undermine coordination. Poor coordination leads to numerous problems including misunderstandings, help requests, lower job satisfaction among team members, and more ambiguous tasks. These problems, in turn, reduced overall project success and prompted professionals to alter their software development process (in this case, from Scrum to Kanban). 

\section{Discussion} \label{sec:disussion}

The initial goal of this research was to investigate how working from home during the pandemic affected software professionals. Our investigation lead to an important determinant of overall project outcomes: team coordination. 




\subsection{Integration with Previous Literature}

\begin{figure}[ht]
  \centering
  \includegraphics[width=0.95\linewidth]{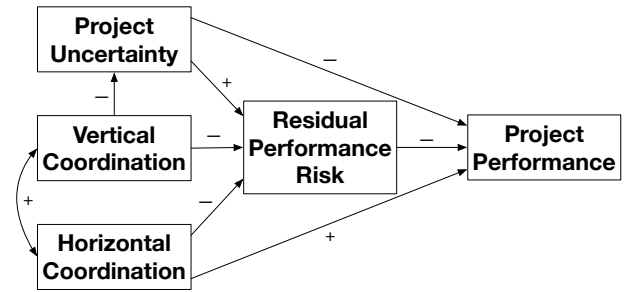}
  \caption{Nidumolu's Coordination Theory (from \cite{nidumolu1995effect})}
  \label{fig:nidumolu}
\end{figure}

Ours is not the first theory of coordination in software engineering. Nidumolu \cite{nidumolu1995effect} proposed a theory distinguishing between vertical and horizontal communication (Fig. \ref{fig:nidumolu}); that is ``the extent to which coordination between users and Information Systems staff is undertaken through vertical means such as authorized entities'' versus ``mutual adjustments and communications, whether through personal or group means'' \cite[p. 194-5]{nidumolu1995effect}. 

Like Nidumolu, we posit that better coordination leads to better project outcomes. We extend Nidumolu's model with several antecedents of coordination (our protective and risk factors) as well as specific problems mediating the relationship between coordination and performance. We do not distinguish between vertical and horizontal coordination because we focused on coordination \textit{within} teams, which is predominately horizontal, at least in agile teams.

Meanwhile, Herbsleb has worked on coordination with numerous colleagues~\cite[e.g.][]{herbsleb2003formulation,herbsleb2007global,herbsleb2006collaboration}. For example, Herbsleb and Mockus model a software project as a series of decisions and define ``coordination'' as the process of ``assigning responsibility for decisions to different people'' \cite{herbsleb2003formulation}. Their perspective is fundamentally different from ours. In our view, decision-making is a poor lens for understanding software development \cite[cf.][]{ralph2016characteristics} and coordination is far broader than assigning tasks (see Section \ref{sec:coordination}). Similarly, Herbsleb and Roberts \cite{herbsleb2006collaboration} model coordination ``as a distributed constraint satisfaction problem,'' which is not wrong, but is quite different from what we mean by coordination. In our view, coordination is about not only determining who will do what but also managing the complex interdependencies among tasks. 

\subsection{Recommendations} \label{sec:implications}

While our research was conducted in a specific, extreme situation, its implications appear transferable to a broader context. The software industry's culture is plainly shifting toward accommodating more flexible work arrangements. This means that remote-first and hybrid teams will be increasingly common. 

Our proposed theory can guide both managers and members of remote-first and hybrid teams on several issues including observing and recognizing coordination problems, and understanding factors that can affect software team coordination post-pandemic. In addition, based on our findings, we can propose the following recommendations to practitioners: 

\subsubsection{Improve communication to cope with the remote context}

Agile methods rely on simple and effective communication. Without routine, face-to-face communication, remote-first teams risk back-sliding into ineffective, documentation-centric knowledge management strategies. Instead, they should adapt their communication mechanisms to allow for more continuous interactions. Including the whole team on important communications is crucial to prevent knowledge silos. In our study we observed three team behaviours related to communication improvement in the remote context: 
\begin{enumerate}
    \item an increase in the number of communication channels that the teams were using to discuss information about the project; 
    \item the use of automated mechanisms to keep everyone updated (e.g. tickets, notes and reminders that are triggered in specific times);
    \item keeping minutes for official agile ceremonies (e.g. planning meeting, product backlog meeting) to make information accessible to everyone.
\end{enumerate}  

\subsubsection{Increase team engagement to support cohesion}

Transitioning from a co-located environment to a home-working environment can rupture team cohesion. As the distance among professionals grows, their trust in each others' work tends to decrease. This scenario will push professionals to seek unnecessary validation, even for uncomplicated activities. We observed that teams attempting to increase team cohesion, engagement and trust in several ways, including creating thematic meetings to add elements of recreation to activities (e.g. open camera and use of costumes; use of a different idiom during the day; virtual happy hours). 

\subsubsection{Create strategies to support parents}

Professionals with family responsibilities (e.g. parenting) are more susceptible to problems triggered by poor coordination. Therefore, it is important to create and maintain strategies to support these professionals on remote activities. This support can be granted by increasing flexibility and enlarging the options to keep updated on the project status (e.g. asynchronous mechanisms).  

\subsubsection{Prepare for method engineering}

Ending up with an unsuccessful project would be the most radical effect of poor coordination. In fact, this scenario would be very unlikely because coordination would only interfere with a number of success criteria, such as, deadline and team satisfaction. However, practitioners need to be prepared for changing process and adapt methods towards overcoming coordination issues that were not contained right in the beginning.

\subsection{Limitations}  \label{sec:limitations}

The above recommendations should be considered in light of the limitations of our study. Grounded Theory should be assessed using ``common qualitative criteria such as credibility, resonance, usefulness and the degree to which results extend our cumulative knowledge'' while ``quantitative quality criteria such as internal validity, construct validity, replicability, generalizability and reliability typically do not apply'' \cite{ralph2020acm}. To maximize credibility, we illustrate the chain of evidence from observations to findings (Table \ref{tab:chainOfEvidence}). To improve resonance, we constantly verified and refined our ideas with interviewees, while cross-referencing interviewee's statements against our observations and field notes. To illustrate usefulness, we provide numerous recommendations in Section \ref{sec:implications}. Finally, we facilitate direct contribution to the cumulative body of knowledge by presenting our findings as a mature, parsimonious theory of software team coordination. 

That said, like most qualitative research, our study does not support statistical generalization from a sample to a population. We study a single site in depth, to generate knowledge that is not accessible through broader but shallower approaches (e.g. questionnaire studies). The resulting theory should be \textit{transferable} to other, similar contexts, and we facilitate transferability by providing as much detail of the site as space allows. Furthermore, while our observations suggest causal relationships between, for example, group cohesion and software team coordination, this is not a controlled experiment through which precedence and covariance can be established while eliminating third-variable explanations. Indeed, many of the phenomena we find important resist experimental closure; that is, they cannot be isolated and manipulated under experimental conditions. Group cohesion, for example, is not a plausible independent variable for a randomized controlled study because we cannot force one group to be more cohesive than another. 

\subsection{Future Work} \label{sec:futureWork}

The limitations discussed in the previous section prompt several potentially fruitful avenues for future work. For example, although our participants have experience with standardized practices of software development and create products for international clients, our findings are context-dependent. Our proposed theory of coordination therefore needs broader validation. It can be transferred and re-assessed in different contexts, particularly post-pandemic remote and hybrid software teams, using case studies. Furthermore, the proposed theory can be transformed into a nomological network---a set of causal propositions relating constructs that can be operationalized using metrics and scales \cite{ralph2018construct}---to support broad validation via questionnaire survey. Indeed, widely used and validated scales for many of the variables observed in this study (e.g. group cohesion, communication effectiveness, trust) already exist. Such transformation entails simply replacing processes (e.g. parenting) and categories (e.g. ill-defined tasks) with variables (e.g. number of children) or constructs (task ambiguity). 


\section{Conclusion} \label{sec:conclusion}
In summary, we conducted a year-long, participant observation, constructivist grounded theory study to investigate how working from home due to COVID-19 affected software teams at a large, South American software development company. The core category that emerged from our study is software team coordination, and the primary contribution of our study is therefore a grounded theory of coordination within software teams (Fig. \ref{fig:theory}). 

This theory identifies several factors that influence coordination (group cohesion, communication, trust, parenting), and suggests that poor coordination leads to problems (misunderstandings, low job satisfaction, help requests, ill-defined tasks) that undermine overall project success, especially from the perspective of team members. However, these problems also trigger method engineering, as professionals adjust their ways of working to avoid impending failure.

While software organizations prepare for more flexible work arrangements, software development tasks will increasingly be performed by hybrid teams of different configurations, some remote, some onsite, some flex (working from home some days and coming to the office other days). Effectively coordinating work in these increasingly complicated configurations will be an important determinant of the overall software project and product success.

Finally, we worry that the trend toward working from home and hybrid teams may lead the software industry to backslide toward pre-agile processes. While much work has attempted to adapt agile methods to global software development, Aagarfalk and FitzGerald's admonishment that ``agile methods and global software development appear to be largely incommensurable'' \cite{aagerfalk2006old} weighs heavily on our thoughts. As we observed in this study, teams compromised many agile practices as they acclimatized to working from home. No one stands up for a stand-up meeting alone in their home office. Working-from-home seems to encourage professionals to focus more on processes, tools, documentation and planning---the opposite of the Agile Manifesto's recommendations \cite{fowler2001agile}. It is therefore incumbent upon us as researchers to make sense of how the emerging trend toward more flexible work arrangements intersects with the cumulative body of knowledge surrounding software processes and team dynamics. The theory we propose in this paper is just a first step on this crucial journey.

\section*{Data Availability}
\label{sec:DataAvailability}

Our screening instrument, interview guide and examples of raw data are publicly available at \url{https://figshare.com/account/home#/projects/132197}. We cannot release interview transcripts, field notes or other communications due to privacy concerns and the risk of re-identifying the participants and projects involved in the research.

\balance

\bibliographystyle{ACM-Reference-Format}
\bibliography{bib}

\end{document}